# Effect of substrate roughness and material selection on the microstructure of sputtering deposited boron carbide thin films

Chung-Chuan Lai[1,*], Robert Boyd[2], Per-Olof Svensson[1], Carina Höglund[1,3], Linda Robinson[1], Jens Birch[2], Richard Hall-Wilton[1,4]

[1] Detector Group, European Spallation Source ERIC (ESS), 224 84 Lund, Sweden

[2] Thin Film Physics Division, Linköping University, 581 83 Linköping, Sweden

[3] Impact Coatings AB, 582 16 Linköping, Sweden

[4] Università degli Studi di Milano-Bicocca, Piazza della Scienza 3, 201 26 Milano, Italy

[*] Corresponding author: Chung-Chuan Lai; E-mail: *chung-chuan.lai@ess.eu*; Phone: +46 721 79 22 68; Post: European Spallation Source ERIC, P.O Box 176, SE-221 00 Lund, Sweden

Abstract

Amorphous boron carbide ($B_4C$) thin films are by far the most popular form for the neutron converting layers in the $^{10}B$-based neutron detectors, which are a rising trend in detector technologies in response to the increasing scarcity and price of $^3He$, the standard material for neutron detection. The microstructure of the $B_4C$ films is closely related to the important properties, e.g. density and adhesion, for the converting layers, which eventually affect the detection efficiency and the long-term stability of the detectors. To study the influence from substrates of different roughness and materials, the $B_4C$ films were deposited on polished Si substrates with Al, Ti, and Cu buffer layers and unpolished Si, Al, Ti, and Cu substrates by direct current magnetron sputtering at a substrate temperature of 623 K. The tapered columnar grains and nodular defects, generally observed in SEM images, indicated a strong shadowing effect where voids were introduced around the grains. The change in the grain size did not show a direct dependence to the substrate roughness, acquired from the surface profile, nor to the mass density of the films, obtained from reflectivity patterns. However, films with non-uniform size of columnar grains were deposited on substrates with high skewness, leading to a drop of mass density from ~95 % down to ~70 % of tabulated bulk density. On the other hand, similar microstructures and mass density were obtained from the films deposited on Al, Ti, and Cu of different roughness and good adhesion were observed from cross-cut adhesion tests, showing the reliability of sputtering deposited $B_4C$ films on common structural materials in neutron detectors.

# 1. Introduction

The under-construction European Spallation Source ERIC (ESS) in Lund, Sweden, is an international project to deliver and operate a powerful neutron source to bolster material science research[1–3]. To take full advantage of the high spallation neutron flux, frontier technologies are being developed to be applied to ESS neutron instruments[1], especially to the neutron detectors of the beamlines. An example is the Multi-Grid detector technology originated from the collaboration with Institut Laue-Langevin, ILL[4–7], for two large area detectors of ESS spectrometry beamlines, C-SPEC and T-REX[8]. In Multi-Grid detectors, isotope-enriched $^{10}B_4C$ thin films are coated on Al substrates as the neutron converting layers, while the detection efficiency can be tuned by varying the number of Al substrates and the thickness-position relation of the $B_4C$ coatings to optimize for, e.g. different wavelengths[9,10]. The design and performance of the Multi-Grid detectors have been published elsewhere[11,12].

One primary advantage of the Multi-Grid detectors is a high saturation limit to cope with higher neutron flux as compared to $^3$He-based detectors, which is to date the standard for neutron detection[13]. Moreover, the $^{10}$B-based technologies, such as the Multi-Grid detectors, are becoming the more economical option as the scarcity of $^3$He gas has increased its price by more than a factor of 20 since 2009[14–17]. Boron carbide has also general advantages over other common neutron converting materials for $^3$He substitution, such as gaseous $^{10}BF_3$, which is toxic, and solid Gd and $^6$LiF, which are chemically less stable.

With regards to the aforementioned benefits, studies of $B_4C$ coatings for neutron detectors has been carried out at ESS Detector Coatings Workshop, in collaboration with Linköping University, since 2010[18]. The coatings, prepared by magnetron sputtering deposition, have been evaluated to uphold the standard for the detector applications, for example, to ensure a high $B_4C$ purity (>95 %) and $^{10}$B-enrichment (>96 %) in the films that are radiation-hard and low in $^{10}$B-depletion under the experiment condition of ESS[19,20]. A screening over various deposition parameters within 0.3 – 0.8 Pa of Ar pressure and 373 – 673 K of substrate temperature has also been done to enable a deposition process at a lower temperature with reasonable residual stress in the films, which is closely related to film adhesion to the substrates[21].

Alongside the development of $B_4C$ deposition processes, the thin film-based detector technologies are also blooming. New technologies can be realized both by optimizing existing $^3$He-based or high energy physics motivated systems, like $^{10}B_4C$-coated gas electron multipliers (GEM)[22,23], or by new concepts that are designed for solid neutron converters, like Multi-Grid, the Jalousie detectors[24,25], and the Multi-Blade detectors[26,27]. Instead of multiple $^{10}B_4C$ layers in the Multi-Grid detectors, the detection efficiency in the Multi-Blade detectors is enhanced by grazing incidence of neutrons onto a single but thicker (>7.5 μm) layer of $^{10}B_4C$ on Ti *blades*. In this way, the effective travelling path of the incident neutrons in the converting layer is longer, while the shortest escape distance for the products from the neutron capturing ($Li^+$ and α particles) is not significantly increased. Other ideas such as depositing $^{10}B_4C$ on to Si-based semiconductor have also been published for the fabrication of light-weight, compact neutron detectors that can easily be connected to read-out electronics[28].

Previously, studies of sputtered $B_4C$ thin films for neutron detector have been focusing on the relationship between the deposition parameters, e.g. substrate temperature and working pressure, and the properties of the films[19,21,29,30], which provide strong foundation to tune in the deposition processes. However, little has been done on different substrate selections other than polished Si wafer, while in some semiconductor-based neutron detectors, metal electrodes are pre-deposited on the surface before deposition of the converting $B_4C$ layer. Moreover, for medium- to large-size neutron detectors, such as aforementioned Multi-Grid and Multi-Blade detectors, the $B_4C$ layers are often deposited directly on metal structures for the consideration of mechanical strength and financial cost.

The surfaces of polished Si wafers and of rolled metal plates can have very different geometrical characteristics, for example that the surface roughness can differ from nm- to µm-range. The effect of substrate roughness can lead to the change in the microstructure of the thin film deposited on top[31], and can further change the properties and the adhesion of the thin film, which is crucial to the performance and long-term stability of neutron detectors. Therefore, a systematic study is needed to explore the effects of different substrate materials and surface characteristics on the properties of the $B_4C$ thin films.

Here, we have prepared $B_4C$ thin films on polished Si substrates, with- or without-metal buffer layers of Al, Ti, and Cu, on unpolished substrates of Si, Al, Ti, and Cu, and on Ti substrates sanded to two different roughness by direct current magnetron sputtering (DCMS). The metals were selected as they are the most widely used materials in practice for building neutron detectors due to their low neutron reaction cross-section. Samples with double buffer layers have also been prepared to reduce the very active surface interaction at the Cu/Si interface. A comparison is made between the $B_4C$ films deposited on relatively smooth Si-based substrates and on the rough metal plates for the microstructure in the electron micrographs and for the surface characteristics acquired from the surface profile. The change in the microstructure of the films on the relatively smooth substrates is related to the change in their mass density, which is a key property directly related to the density of $^{10}B$ atoms in the film and hence the neutron detection efficiency. Finally, the microstructures of the films on defective Si substrate surfaces are presented, showing that surface characteristics, like protruding features from the surface, can cause higher density of voids in between the columnar grains and lower the mass density of the $B_4C$ films.

## 2. Experimental Details

### 2.1. Sample preparation

The B$_4$C thin film samples prepared and analyzed in this project are summarized in Table I. The samples are sorted in two groups to aid later discussion. In the *Polished* sample group, six B$_4$C films were deposited on mirror-polished Si(100) pieces (525 µm thick, *p*-type), of which five samples were deposited with a metal buffer layer prior to the B$_4$C deposition. The *Unpolished* group consists of four B$_4$C films deposited directly on the backside (etched surface) of a Si(100) piece and on rolled Al (Al-5754), Ti (grade 2), and Cu (ETP grade) plates, all with a thickness ~0.5 mm. Lastly, the two samples in the *Sanded* group are B$_4$C films deposited on rolled Ti plates (same kind with *Unpolished* group), while the surface of the plates have been finished by hand-polishing with P320 and P2000 sandpapers. All substrates were cleaned sequentially in 5 minutes ultrasonic bath of acetone and isopropanol, and blow-dried with compressed N$_2$ gas prior to the first depositions.

**Table I.** Summary of the B$_4$C thin film samples in this project.

| Sample group | Substrate Material | Surface treatment | Intended buffer layer |
|---|---|---|---|
| *Polished* | Si | Mirror-polished | - |
| *Polished* | Si | Mirror-polished | Al |
| *Polished* | Si | Mirror-polished | Ti |
| *Polished* | Si | Mirror-polished | Cu |
| *Polished* | Si | Mirror-polished | Cu/Al |
| *Polished* | Si | Mirror-polished | Cu/Ti |
| *Unpolished* | Si | Etched (backside of wafer) | - |
| *Unpolished* | Al | Rolled (shelf material) | - |
| *Unpolished* | Ti | Rolled (shelf material) | - |
| *Unpolished* | Cu | Rolled (shelf material) | - |
| *Sanded* | Ti | Rolled and sanded to P320 | - |
| *Sanded* | Ti | Rolled and sanded to P2000 | - |

**Table II.** Summary of the process parameters for the depositions of different materials.

| Target material (purity in wt.%) | Ar pressure (Pa) | Substrate temperature (°C / K) | Cathode power (W) | Substrate potential |
|---|---|---|---|---|
| Al (99.9) | 0.8 | 350 / 623 | 8000 | Floating |
| Ti (99.9) | 0.2 | 350 / 623 | 9500 | Floating |
| Cu (99.9) | 0.3 | 80 / 353 | 4000 | Floating |
| $B_4C$ (99.9) | 0.2 | 350 / 623 | 4000 | Floating |

The $B_4C$ films and the buffer layers were prepared by DCMS deposition with a commercial CemeCon CC800/9 deposition system in ESS Detector Coatings Workshop in Linköping, Sweden. The system was equipped with four sputtering targets with surface area approximately 88*500 mm$^2$, facing sideways towards the substrates, and was evacuated to a base pressure of ~1.5*10$^{-4}$ Pa before the depositions. The samples were mounted on a rotating sample table (~640 mm in the diameter, ~1 rpm in the speed) which provided a horizontal motion to ensure a better heating and deposition uniformity. Table II summarizes related parameters for Al, Ti, Cu, and $B_4C$ depositions. After each deposition, the system was cooled down to the room temperature before venting to air to avoid oxidation of the samples at elevated temperature.

*2.2. Sample characterization*

Images of cross-sectional microstructure and surface morphology of the samples were acquired with scanning electron microscopy (SEM) in a Zeiss LEO 1550 Gemini system, except the cross-sectional images of the metal substrates. Energy dispersive X-ray spectroscopy (EDS) mapping of chemical elements at interfaces was done for the samples with a Cu buffer layer due to the known high Cu diffusion rate in Si. The mapping was collected by SEM attached detector (Oxford Instruments) and analyzed by Aztec program from the same supplier.

The films deposited on metal substrates in *Unpolished* group were treated in a focused ion-beam SEM system (FIB-SEM, Zeiss Crossbeam 1540 EsB) for cross-sectioning of the samples to avoid mechanical cutting and polishing of soft metal substrates, which can alter and mask the deposited film. A Pt protective layer was deposited on the surface of the samples before sputtering of by use of a 30 kV Ga ion-beam. After cross-sectioning, the images of the samples with metal substrates were acquired directly in the same FIB-SEM system. For imaging the sample normal was tilted 54° to the electron beam in order to view the cross section. A tilt correction of 36° was used to compensate.

Surface profile of the samples before and after deposition of the B$_4$C films were measured by a Bruker DektakXT stylus profiler equipped with a 2 μm stylus. The traveling speed of the stylus was fixed to 200 μm/s and the force to 3 μN. Since the rolling process for the metal substrates is directional, the surface profiles of the rolled substrates were recorded perpendicularly to the rolling direction to measure the surface features introduced by the rolling process. The profiles were further analyzed to obtain surface roughness ($R_{rms}$, root-mean-squared roughness) and profile characteristics ($R_{sk}$ and $R_{ku}$, skewness and kurtosis, respectively) using Vision64 program from the same supplier. In order to obtain more standardized results from the analysis, selection of the long and short cutoff wavelength as well as the required length for evaluation was following the suggestion from ISO4288:1996 standard[32]. For each surface, 10 to 15 sites were measured, i.e. 10 to 15 profiles recorded, and the mean value of $R_{rms}$, $R_{sk}$, and $R_{ku}$ from all profiles was presented.

The samples for collecting the surface profile of the Cu layers additionally annealed in the same vacuum system at 350°C for 3 h prior to the measurements in order to create a similar surface where the B$_4$C layers were later deposited on. This is due to that the Cu-layers were deposited at a lower the substrate temperature (~80°C) compared to the other metal and B$_4$C layers.

Mass density of the B$_4$C thin films on polished Si substrates were studied by X-ray reflectivity (XRR) performed in a Malvern Panalytical Empyrean X-ray Diffractometer with a Cu light source. The collected XRR patterns were analyzed and fitted for the critical angles in GenX (version 3.5.0) [33]. The mass density of the samples in *Unpolished* group and in *Sanded* group, however, could not be acquired due to high surface roughness that influences the shape of the XRR patterns. The XRD pattern of the $B_4C$ film on Si substrate were acquired in the same instrument to confirm that the $B_4C$ phase in the sample is mostly existing in X-ray amorphous-form. Figure 1 gives examples of the fitted XRR patterns and of the collected XRD pattern.

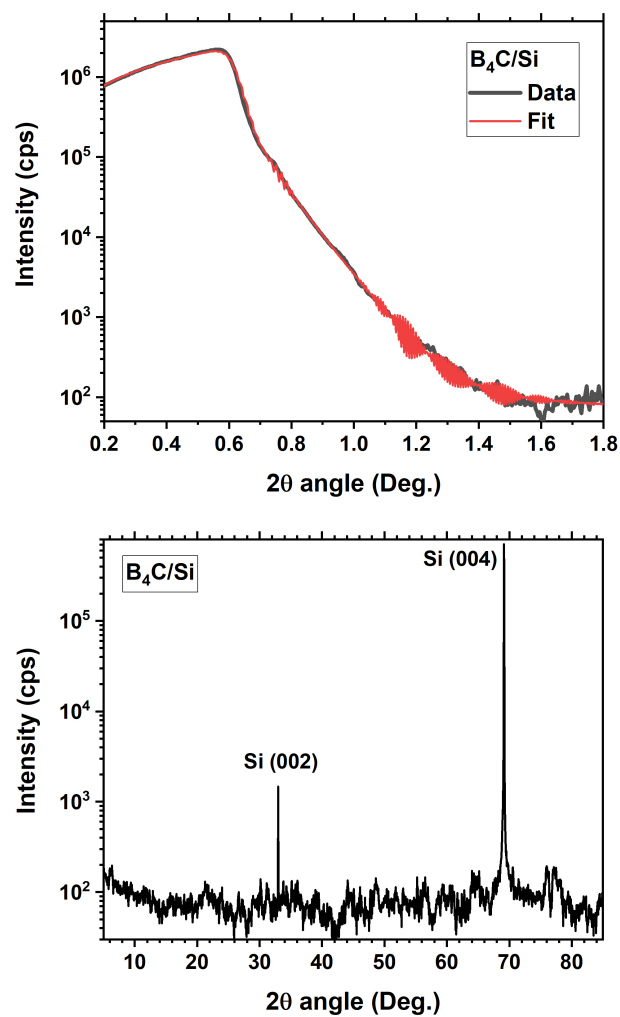

**Figure 1.** [Colored] Examples of (top) a fitted XRR pattern and (bottom) an XRD pattern of the $B_4C$-coated Si sample. The diffraction peaks from the Si substrate are labelled.

*2.3. Adhesion test*

The adhesion of the B$_4$C films were examined by cross-cut tests, where, for each sample, 7 parallel cuts were made with a sharp blade on the surface with 1 mm apart followed by another 7 parallel cuts made perpendicularly on top of the first 7 cuts. See also Figure 2 for an example of cross-cuts made on the B$_4$C-coated rolled Al sample. All the cuts were made through to the substrate without cutting repeatedly on the same location, and the debris from the cuts were blown away with compressed N$_2$ gas. A section of fresh tape (Scotch Crystal Tape, 3M) was put on the cross-cuts, pressed over uniformly with a pencil eraser, and let relaxed for about 30 s before pulled off quickly at a reversed direction as parallel to the surface as possible. After removing the tape, the intactness of the coatings at the cross-cut area was examined and graded from 0 (no coatings remained) to 5 (coatings fully remained) with a linear scaling relation in between to how much area of coatings remained on the sample, e.g. grade 2 mean ~40 % area remained.

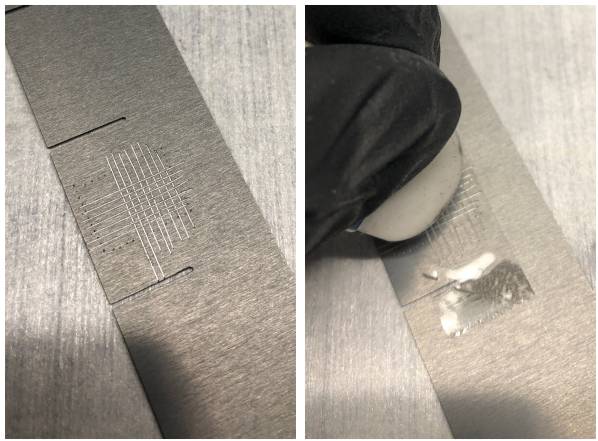

**Figure 2.** [Colored] Example photos of (left) the cross-cuts made on the B$_4$C-coated rolled Al sample and (right) the tape over the cross-cut area. The tape was pressed by a pencil eraser to get a better surface contact and uniform pressure distribution.

## 3. Results

*3.1. Microstructure and surface morphology*

SEM micrographs are presented in Figure 3 for the microstructure and the surface morphology of the $B_4C$ thin films in the *Polished* group. The thickness was determined from the cross-sectional images to be ~1.3 µm in general for the $B_4C$ films, and ~1.3 µm and ~1.0 µm for the Al and Ti layers, respectively. The thicknesses of individual layers observed in the micrographs are in good agreement with the intended thicknesses for $B_4C$ = 1.2 µm, Al = 1.0 µm, Ti = 1.0 µm. The actual thickness of the Cu layers cannot be determined easily due to undefined layer boundaries with adjacent layers, despite an intended thickness of 1.0 µm, the same with other metal layers. The intermixing of Cu and other layers will be discussed later.

The $B_4C$ films primarily show tapered columnar structure in the growth direction, i.e. normal to the substrate surface, that ends with the "cauliflower-shaped" surface shown in the corresponding planar-view images[19]. The diameter of the columns varies from few tens of nm, as deposited without a buffer layer, up to the range of sub-micrometer or micrometer, as in the samples with a Cu/Al or Cu/Ti double buffer layer. However, regardless of the column width, pits and trenches in different sizes (with darker contrast) are found around the columns. This can indicate a growth condition with a limited adatom mobility for the film-forming particles, where geometrical shadow effect plays important role. It can be compared to the extreme shadowing cases demonstrated by Mukherjee et al.[34] that when homologous temperature $\theta$ (the substrate temperature to melting temperature ratio, both in Kelvin) is as low as <0.3, large voids are created around the taller grains due to blocking of the incoming particles to land and lack of diffusion of the adatoms.

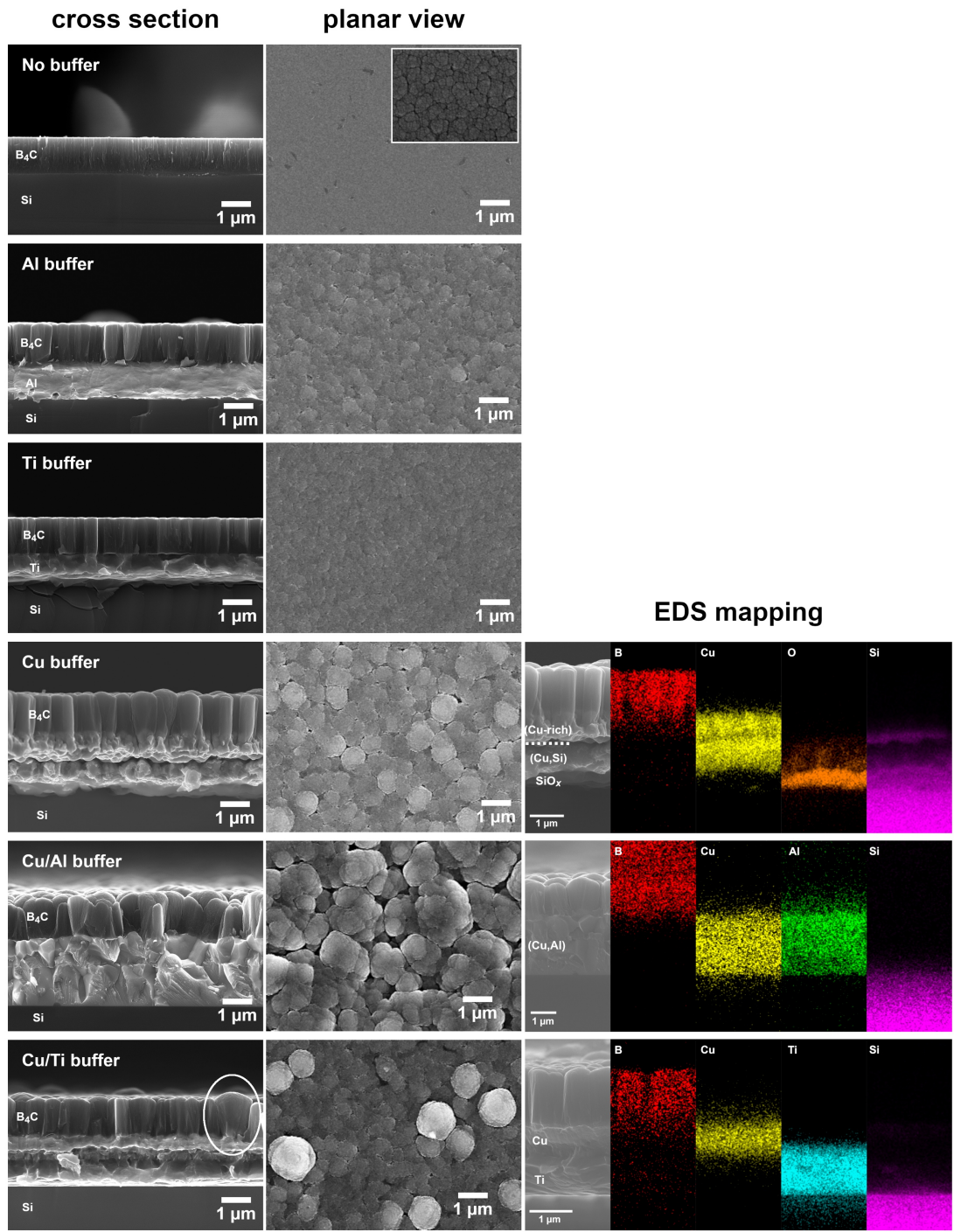

**Figure 3.** [Colored] (Left column) Cross-sectional and (middle column) planar view SEM images of samples in the *Polished* sample group. All images are presented with the same magnification except for the inset. Labelling of the layers in the images was done according to the EDS mapping results instead of the intended buffer layers listed in Table I. (Right column) EDS mapping of B, Si, and corresponding chemical elements in the buffer layers of the Cu-containing samples.

The height of the columnar grains is not distributed uniformly on all samples in the group. On those samples with no or a single buffer layer, most B$_4$C columns are rather uniform in both height and width, with only a few taller protrusions scattered over the surface, while on the samples with double buffer layers, the height and width differences among the grains increase significantly across the surface. A taller grain is highlighted in the blue circle in Figure 3 in the film deposited on a Cu/Ti double layer and the tall grains of similar kind can be related to the large grains in brighter contrast in the planar view image. In addition to the protruding columns, the film grown on a Cu/Al double layer contains high density of deep and wide pits and trenches in between the columns, with smaller grains scattered over the surface.

We have observed that the Cu/Si interface is rough and undefined in the cross-sectional image, in contrast with the Al/Si or Ti/Si interfaces. It is well-known in the semiconductor industry that a Cu/Si contact can react to form intermetallic compounds, primarily Cu$_3$Si-like, already at a temperature around 100°C (373 K)[35,36]. The substrate heating for the B$_4$C deposition in this work, 350°C, is higher than the reported reaction temperature. The The EDS mapping (right column in Figure 3) has confirmed the existence of an intermixing layer of Cu and Si in between a Cu-rich and a silicon oxide, SiO$_x$, layer. The intermixing is likely initiated by an outward diffusion of Cu into the relatively immobile Si matrix at such temperatures[37], leaving a Cu-rich layer on the opposite side of the buffer layer. The formation of the oxide layer is expected due to the room temperature oxidation of Cu$_3$Si-like phase when the samples were removed from vacuum into the air after the Cu deposition, while the reannealing further enhance the formation of silicon oxides at the expense of copper oxides[38]. As a result, the EDS signal of Cu was collected in an extended layer of ~2 μm compared to the intended thickness of 1.0 μm for the deposition. In addition to the reacted layers, a crack, highlighted by the dotted line, was found in between the Cu-rich layer and the intermixing layer. The crack can be introduced while the cleaving of the cross-sectional SEM sample and possibly shows weak adhesion of the deposited film to the substrate.

An indication of intermixing can also be observed at the Cu/Al and Cu/Ti interfaces in the EDS mapping. The results match with the reported experiments done in searching of diffusion barriers for Cu/Si interfaces in a similar temperature range[39,40]. The Cu/Al interface has a nearly complete intermixing between the two layers, possibly due to its relatively low eutectic temperature at around 823 K[41], which further leads to a complete penetration of both elements into the counter layers and even Cu penetration into the Si substrate. Crystalline facets can also be seen in the micrograph, indicating the formation of an intermetallic phase with low melting temperature. Unlike the Cu/Si case, no crack or void is seen at the interfaces after the cleaving, showing an acceptable film adhesion. The Cu/Ti interface has the shortest intermixing range under the EDS mapping compared to the other two cases, but it is difficult to exclude the possibility of an overlap in the interaction volume, which is typically about 1 – 2 μm in diameter for the given electron-beam energy. In both cases of double buffer layers, the final Cu surfaces are rough compared to the case without a buffer layer or with a single buffer layer of Al and Ti. The cross-sectional and planar-view SEM images of the *Unpolished* group samples are presented in Figure 4. The $B_4C$ films in this group are generally uniform in the thickness that follows the surface contour of the substrates, despite that the initial surface roughness on the substrates are higher than for the *Polished* group. The microstructure of the $B_4C$ film on the etched Si substrate is very similar to the polished Si substrate, where both have fine fibrous structure with individual fibers having a width in the 10 - 100 nm range. On the other hand, the microstructure of the films on the rolled metal plates, is comparable with the $B_4C$ films grown on an Al or Ti single layer in Figure 3. The size of the cauliflower-shape features is of a similar sub-micron size and a low number of voids and pits are found in between the grains, except the ones that originate from the features of the metal substrates.

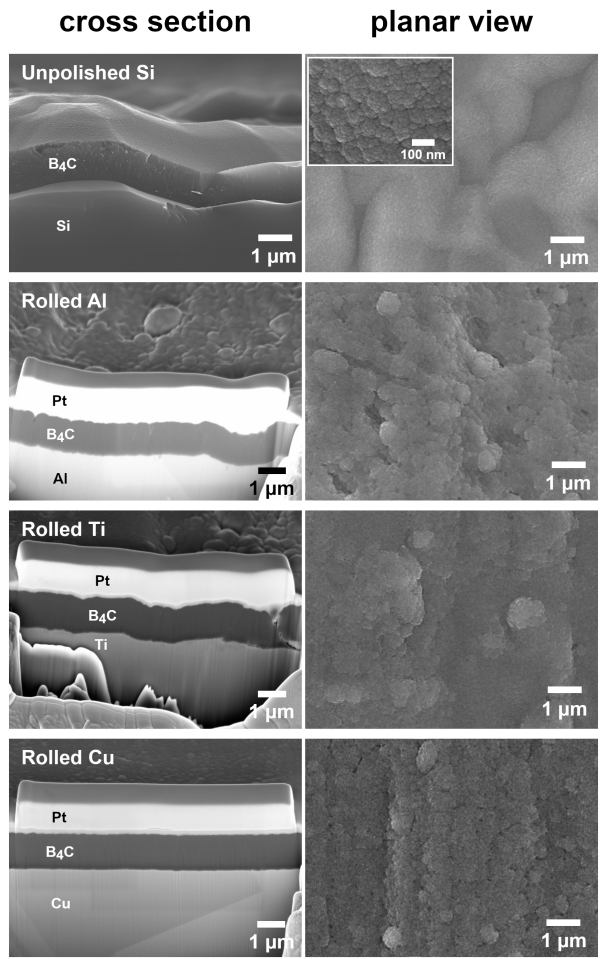

**Figure 4.** (Left column) Cross-sectional and (middle column) planar view SEM images of samples in *Unpolished* group. All images are presented with the same magnification except for the inset. The inset shows a planar view SEM image of unpolished Si taken at a higher magnification.

In the $B_4C$ films of both sample groups, features with an inverted cone shape were found embedded in the films, as shown in Figure 5(a) and (b), on the polished Si substrate, and Figure 5(c), on the rolled Al plate. These features can be compared to the nodular defects from literature of thin films deposited with physical vapor depositions[42] and have also been reported in thin films of sputtering deposited $B_4C$[29]. Nodular defects are usually observed surrounding by voids and therefore are loosely adhered to the bulk region of the coating and the substrate, as highlighted in red circle in Figure 5(b).

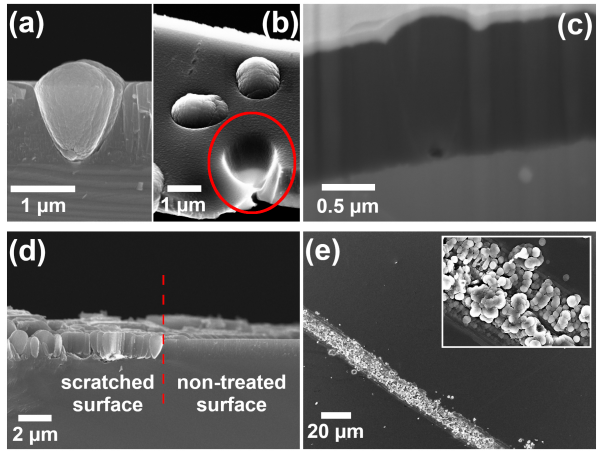

**Figure 5.** [Colored] SEM images of defects in the B$_4$C films. Nodular defects intergrown with the B$_4$C film on (a)&(b) the polished Si substrate and (c) the rolled Al plate. Notice that (a)&(b) are from a cleaved sample and (c) from FIB preparation. High density of nodular defects concentrated at a scratch on the Si surface introduced prior to the B$_4$C deposition (d) in cross-section and (e) planar view. The inset of (e) shows a planar view image over the scratched area at a higher magnification.

To study the effect of substrate roughness to the film structures and the formation of nodular defects, we intentionally introduced some scratches onto a polished Si substrate with a steel scriber prior to the deposition. In the vicinity of the scratches, we observed an abrupt transition of microstructure in the deposited B$_4$C film. In Figure 5(d), the surfaces scratched and non-treated are separated by the red dotted line, where the B$_4$C film on the latter is similar to the film without a buffer layer in Figure 3. The film on the scratches, however, grown in high concentration of nodular defects, surrounded by high number of voids, and with a much higher film roughness. Figure 5(e) shows a contrast in planar view where the defects follow the scratch in a straight line and the rest areas are relatively free from the defects. In between the nodular defects some deep trenches can be seen, which is similar to the film on the Cu/Al double buffer layers.

## 3.2. Roughness and height characteristics

Figure 6 summarizes the mean values of $R_{rms}$, $R_{sk}$, and $R_{ku}$ extracted from the surface profiles of the samples, where the values before and after the B$_4$C deposition are put next to each other for the ease of comparison. Before the deposition, the polished Si substrates has a low surface roughness (< 1 nm), which increases slightly ($\Delta R_{rms}$ < 2 nm) after the addition of the Al or Ti buffer layer but much more ($\Delta R_{rms}$ > 15 nm) after all cases of the Cu deposition. This is in line with the cross-sectional images in Figure 3, where a rougher interface can be seen under the B$_4$C thin film in the samples planned with a Cu buffer layer. Substrates in *Unpolished* group, on the other hand, have 1 – 2 order higher $R_{rms}$ values (150 – 400 nm) than the ones in *Polished* group due to the finishing process of the substrate manufacturing, e.g. etching for Si or rolling for metal plates.

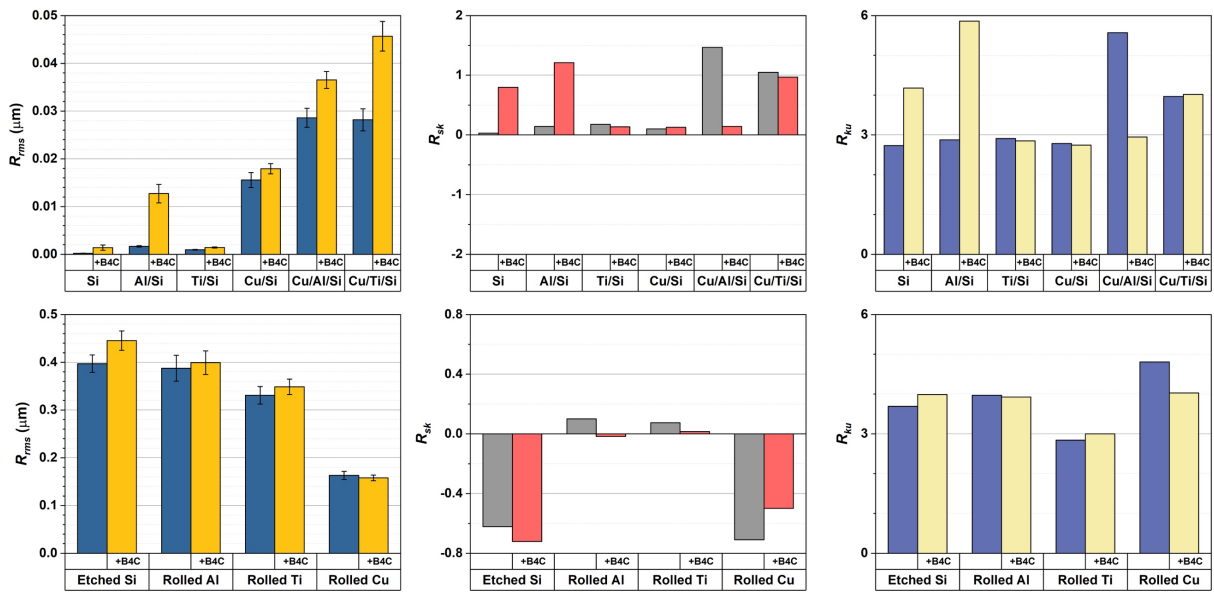

**Figure 6.** [Colored] Mean $R_{rms}$, $R_{sk}$, and $R_{ku}$ values of the samples, before and after the B$_4$C deposition, in (upper row) *Polished* group and (lower row) *Unpolished* group. The error bars in the roughness graphs show the range of the first standard deviation. Notice that for skewness and kurtosis, $R_{sk} = 0$ and $R_{ku} = 3$, are intentionally placed in the center of the graphs.

After the B₄C deposition, we observed an increase in $R_{rms}$ for all samples in both groups except for the rolled Cu substrate. This is likely due to the diffusion-limited growth condition where shadowing effect is dominating, and therefore surface features created, e.g. voids, cannot be smoothened out easily. A positive correlation can be seen between the column width or the size of the cauliflower-shaped feature in Figure 3 and the $R_{rms}$ value of the samples in *Polished* group, showing that the microstructure in these films is mainly responsible for the final surface roughness. Such correlation, however, is not seen in *Unpolished* group, and the increments in $R_{rms}$ are generally in smaller proportion with the value before the B₄C deposition as compared to the samples in the other group. This indicates that, instead of the microstructures in the films, the substrate roughness in *Unpolished* group gives the major contribution to the final surface roughness.

The height characteristics of the samples are assessed by $R_{sk}$ and $R_{ku}$, which tell the population density of the data points in the recorded surface profile. The polarity of the skewness value shows if the tail of the asymmetry in the data distribution appears on the right (positively skewed) or the left (negatively skewed) side of the majority, while the amplitude tells the degree of asymmetry. A positively or negatively skewed surface profile normally indicates a surface with more protruding peaks or more sinking valleys, respectively. The kurtosis value describes the shape of the data distribution with respect to a normal distribution ($R_{ku}$ = 3). When $R_{ku}$ > 3, the data population has a lower density at the tails of the distribution, showing a more concentrated distribution and a surface profile with sharper changes in the amplitude. On the contrary, when $R_{ku}$ < 3, the data distribution is flatter and the surface profile is likely to have more gradual changes in the amplitude.

The polished Si substrate in *Polished* group have the surface profiles rather symmetrically distributed around the mean line, while $R_{sk}$ increases, i.e. the profile becoming more protruding, with increasing number of buffer layers added. The larger deviation of $R_{ku}$ from 3 for the samples with two buffer layers also indicates a more distinctive increase in population density of the outliers than the samples with no or only one buffer layer.

Formation of peak or valley features in the B$_4$C films does not seem to have a direct relation with the roughness or skewness of the underlying surface. After the B$_4$C deposition, a large increase in both $R_{sk}$ and $R_{ku}$ values for the samples with no buffer layer and with an Al buffer layer shows that the film surfaces have grown more into steep protrusions from a relatively feature-less substrate. By looking at the images in Figure 3 and Figure 5(a)&(b), the protrusions in the films are likely the nodular defects that scattered over the surface. However, this increase is not seen in the samples with a Ti or Cu buffer layer even though the surfaces are also quite feature-less. Furthermore, it is exactly in the contrary for the Cu/Al/Si sample, where the skewness and kurtosis drop sharply after adding the B$_4$C layer.

In *Unpolished* group, the substrates are of peak type for rolled Al and rolled Ti and of valley type for etched Si and rolled Cu according to the polarity of the skewness. For the etched Si, the surface commonly becomes valley type after an etching process due to development of etch pits at defective sites, while for the rolled metal plates, the surface is mainly affected by the surface feature of the contacted counterpart during the rolling process. Notice that the scale for the skewness is smaller compared to *Polished* group, indicating to a relatively less outstanding surface features, e.g. shorter peaks and less shallow valleys, from surfaces with already high roughness. The same observation is found on samples in *Sanded* group, shown in Figure 7, where the skewness is in the similar order of *Unpolished* group samples.

Deposition of B$_4$C thin films over the unpolished and the sanded samples does not substantially change the height characteristics, as shown by the small differences of $R_{sk}$ and $R_{ku}$ values in Figure 6 and Figure 7. This is similar to the observation of the small changing in $R_{rms}$ values after coating, where the initial surface features from the manufacturing process are more pronounced than those from the deposited films.

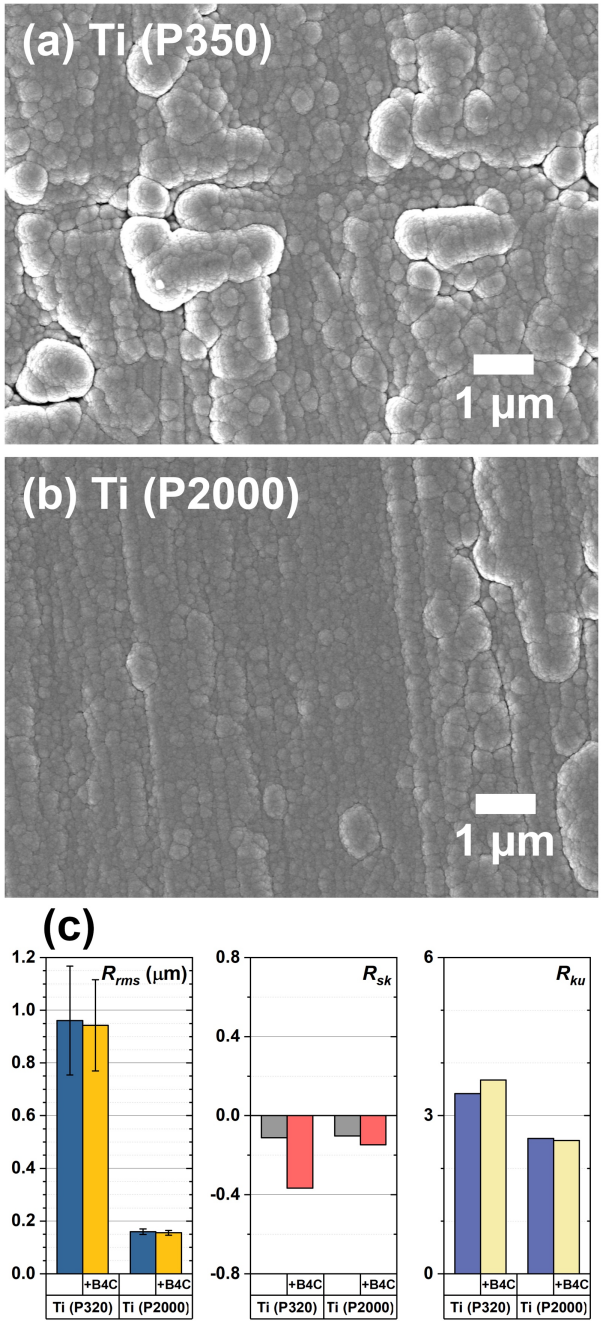

**Figure 7.** [Colored] (a) and (b) Planar view SEM images of the samples in *Sanded* group. (c) Mean $R_{rms}$, $R_{sk}$, and $R_{ku}$ values of the same group, before and after the B$_4$C deposition. The error bars in the roughness graphs show the range of the first standard deviation. $R_{sk} = 0$ and $R_{ku} = 3$, are intentionally placed in the center of the skewness and kurtosis graphs.

The similarity in the height characteristics between the metal plate samples in *Unpolished* and *Sanded* groups reflects to the same features in their surface SEM images in Figure 5 and Figure 7, where cauliflower features with similar size are packed together in most areas and leave small voids in between. However, when $R_{rms}$ increases significantly from ~0.15 μm for Ti (P2000) to ~0.95 μm for Ti (P320), the size of the cauliflower increases. Those larger features create deeper valleys between the grains, which lines up with a change in the kurtosis value from below to above 3, i.e. the number of outlier points in the surface profile has increased.

*3.3. Density measurement*

The mass density of the $B_4C$ films in *Polished* group has been acquired from XRR pattern fitting and summarized in Table III. Films deposited with or without a single buffer layer all have a general density around 2.4 g/cm$^3$, or around 95 % of tabulated mass density (2.52 g/cm$^3$) for bulk $B_4C$. The density of the films is mostly consistent with our previously reported values of between 2.2 and 2.4 g/cm$^3$, from similar deposition conditions[19,21]. A more significant reduction is observed in the density of the $B_4C$ films deposited on a double buffer layer, where the relative density is around 73 % and 93 % of the bulk value for the respective samples on Cu/Al and Cu/Ti buffer layers.

**Table III.** Mass density of the $B_4C$ thin films on the samples in *Polished* group.

| Substrate | Si | Al/Si | Ti/Si | Cu/Si | Cu/Al/Si | Cu/Ti/Si |
|---|---|---|---|---|---|---|
| **$B_4C$ Density (g/cm$^3$)** | 2.38 | 2.36 | 2.40 | 2.38 | 1.83 | 2.34 |
| **Uncertainty max** | +0.070 | +0.074 | +0.063 | +0.079 | +0.045 | +0.015 |
| **Uncertainty min** | -0.003 | -0.076 | -0.043 | -0.009 | -0.037 | -0.017 |

Comparing to the micrographs in Figure 3, the columnar structures in the films on Cu/Al double layer vary widely in both width and height with the deep pits packed in between, which are possibly responsible for the large reduction in the average film density. Notice that the density values reported here are calculated from results of XRR pattern fitting, which express only the *average* electron density in the film over a scanned area by the X-ray beam. In other words, the local density of the features (e.g. the columns) is not known and cannot be directly compared between different grains or samples. The relatively high uncertainty values reported here stem from the XRR fitting done with high uncertainty in the thickness parameters as the $B_4C$ layers and the buffer layers are too thick for a reasonable fitting.

*3.4 Adhesion test of the $B_4C$ films*

The grades from the cross-cut tests of the $B_4C$ films are summarized in Table IV with *Polished* group in the upper rows and the other groups in the lower rows. Almost all samples were graded with 5, i.e. with coatings fully remained after the tapes removed, except for the samples with either a Ti or Cu buffer layer or both. Therefore, the adhesion is generally good for the sputtered $B_4C$ films on Si, Al, Ti, and Cu surfaces from various preparation methods.

**Table IV.** Grades from the cross-cut tests of the $B_4C$ films. The cross symbol (†) after a grade indicates that the film was delaminated from the underlying interface with the Si substrate instead of the interface with the $B_4C$ films.

| **Substrate** | Si | Al/Si | Ti/Si | Cu/Si | Cu/Al/Si | Cu/Ti/Si |
|---|---|---|---|---|---|---|
| **Grade** | 5 | 5 | 0† | 0† | 5 | 2† |
| **Substrate** | Unpolished Si | Rolled Al | Rolled Ti | Rolled Cu | Ti (P320) | Ti (2000) |
| **Grade** | 5 | 5 | 5 | 5 | 5 | 5 |

By inspecting the tapes and the samples from those combinations scored less than 5, we realized that the coatings were actually removed from the metal-to-silicon interface instead of the interface to the $B_4C$ layers. This means that the $B_4C$ layer has a stronger adhesion towards the Ti and Cu layers than the latter to the polished Si wafer, and lines up with the general score 5 on metal plates with different roughness in *Unpolished* and *Sanded* groups.

4. Discussion

The microstructure of the $B_4C$ films shown in Figure 3, e.g. taper columns surrounded by voids, tall protruding columns, and nodular defects, hints to a diffusion limited condition. This is in line with a low substrate temperature and a general lack of ion bombardment in a DCMS process without additional substrate bias during deposition. The structure zone diagram (SZD) can be referred to here to qualitatively evaluate the relation between the thin film microstructure and the assessed deposition parameters. A homologous temperature of $\theta \sim 0.23$ (taking the melting temperature of $B_4C$ phase = 2723 K when carbon content is 18.6 at.%[43]) falls in zone Ic of the structure zone diagram (SZD) proposed by Mahieu et al, or zone T by Barna and Adamik and zone II by Thorton[44–46]. Despite the inconsistency in the terms, the authors all described the range of $0.2 \leq \theta \leq 0.3$ in having low mobility for the film forming adatoms to enable very limited surface diffusion, not across the grain boundaries. The microstructure is mainly columns separated by voids and the growth mechanism is dominated by the self-shadowing of the columns, similar to the observation in this work. Engwall et al. in a recent publication has studied >1 μm thick $B_4C$ films deposited by DCMS at a substrate temperature of 450 °C (723 K), which corresponds to $\theta \sim 0.27$[29]. With a homologous temperature in the same range, they have also observed a very similar microstructure and formation of nodular defects in films deposited without a buffer layer. The microstructure-growth condition relations from the works confirm the empirical prediction from the SZD for synthesis of $B_4C$ films within such temperature range.

Comparing between Figure 3 and Figure 6, we can see a positive relation between the width of the columnar $B_4C$ grains and the roughness of the underlying layers, i.e. the columns become wider with increasing surface roughness. In such conditions with limited diffusion through grain boundaries, the columnar grains cannot grow in width through coalescence but by competitive growth from geometrical shadow effect. The surface roughness of the substrates/buffer layers serves as templates with different height to the initial stage of film growth, so that grains nucleate at higher sites (on peaks) have local advantages to shadow over the ones nucleate at lower sites (in valleys). In this case, the limitation to the width of the columns is set by the density of the advantageous sites over the substrate surface. Therefore, on a smooth substrate like a polish Si-wafer, fibrous fine columns are observed as no apparent site is in favor of shadowing. As the surface roughness increases, the columns grow wider since more sites in between the columns are shadowed.

However, the surface roughness to microstructure relationship does not hold on the substrates of the *Unpolished* group, as shown in Figure 4, where the roughness values are in general 1 to 2 orders higher than for the polished samples. The SEM images of the *Unpolished* group without many voids observed indicates to a film microstructure and density similar to the *Polished* group samples with low surface roughness. The inconsistency of the microstructure change is related to the fact that surface roughness describes mainly the amplitude with respect to the mean height of the profile (after removal of features with longer wavelength, e.g. waviness) but does not contain information of other height and spatial characteristics for the surface features. For example, the $B_4C$ films with larger columns and clear voids were deposited on the double buffer layers (Cu/Al and Cu/Ti) with relatively high skewness and kurtosis values, i.e. with more sharp protruding features on the surfaces. It can be understood intuitively that for a non-grazing incidence of deposition flux, a sharper protrusion can shadow larger areas in close vicinity than a slow-climbing hill, as observed in the case of unpolished Si wafer.

Of the samples of the *Polished* group, the surface roughness and characteristics are likely cumulated over added buffer layers from, e.g. crystalline facets as shown in the Cu/Al double layers, as shown in increasing surface roughness and skewness. Sharp protrusions on the substrates are usually extrinsically originated, such as dusts or flakes coming from coated vacuum parts of a deposition chamber[29] that lands on the substrate surface. This results in randomly distributed nodular defects, as observed in Figure 5(a)-(c). When the area density of the surface features increases, the number density of the nodular defects increases. This is also demonstrated by depositing a $B_4C$ coatings on the scratched Si wafer surface, as shown in Figure 5(d)&(e). One can observed an abrupt transition from fine columnar structure into concentrated nodular defects on the scratch, where the latter resembles the wide columnar grains found in the films on the double buffer layers. The similarity implies that the microstructures have a similar origin from the geometrical shadow effect, as discussed above, in the growth condition with low adatom mobility.

Change of the column width does not seem to have large impact on the mass density of the films, as shown in Table III, as long as the columnar grains are uniform in height and width to minimize the voids in between. Brett et al. has showed, with simulations, that the mass density of the nodular defects is indeed reduced at the grain boundaries but is nearly the same as a bulk film in the central part of the nodules[42]. Hence, the width of the columnar grains, or concentrated nodular defects, changes the local distribution of the grain boundaries but is probably less impactful to the average film density. The inhomogeneity of the grain size, however, introduces large voids into the $B_4C$ film as shown on the Cu/Al double layers and on the scratch over the Si wafer. Consequently, a significant reduction can be seen in the mass density of the former case.

On the other hand, such deep valleys are not seen in the microstructure of the $B_4C$ films on the samples in *Unpolished* group. In fact, the local microstructure of the unpolished samples is more similar to the polished samples with no or a single buffer layer, which implies to a similar mass density in these films despite the large difference in roughness values. The deep valleys appear again in Figure 7(a), where the Ti plate is hand-polished to a surface roughness of $R_{rms}$ ~ 0.95 μm. Notice that this value is at least two-fold compared to those for the off-the-shelf metal plates produced by rolling, which are shown in Figure 6 for Al, Ti, and Cu plates. Comparing Figure 7(a) and (b), further polishing the Ti plate also reduced the voids on the surface without much changing in the skewness of the surface profile. Since most detector applications use $B_4C$ films deposited directly on the rolled metal plates or foils, little to no mechanical treatment is needed for coatings with low density of voids. However, a mechanical grinding or polishing can still be considered to reduce overall roughness, to change the polarity of the skewness, or to remove local defects with high amplitude in the surface profile.

For applications that prefer a direct integration of $B_4C$ films on Si substrates, the interfaces on both polished and unpolished Si are smooth with no indication of intermixing for the given deposition parameters in this work. The $B_4C$ films show generally good adhesion with no stress-introduced voids in the SEM images to the selected substrate materials (Si, Al, Ti, and Cu) commonly used in the neutron detectors. However, the addition of Cu layers to Si substrates can introduced deterioration of Si structures and bad adhesion (at the Cu/Si interface) or high density of protruding features (multiple buffer layers). Using single layer of Al shows less adhesion problem to Si substrate than Ti or Cu, and a film density similar to direct deposition on Si surface.

## 5. Conclusion

We have shown that the $B_4C$ films deposited in a growth condition with a low homologous temperature (~0.23) and without high flux of ion bombardment (DCMS) is dominated by geometrically self-shadowing. This is consistent with other experimental results from literature and also agrees with the general description of thin film microstructure in SZDs, which are published mainly for deposition of thick metal films. With the same deposition parameters, the columnar grains are grown with different widths depending on the features on the substrate surfaces, where sharper protruding features enhance the shadow effect from the initial nucleation stage, leading to growth of wider grains. Since the shape of the surface features also plays an important role to the development of the microstructure, it is not enough to predict final properties of $B_4C$ films with solely the surface roughness. Other surface characteristics, such as skewness and kurtosis, should also be considered. The average mass density of the film, however, is less affected by the size of the grains or the selected substrate materials but more by how well the columnar grains packed together. On an etched or a rolled metal surface, where protruding features are not so pronounced, a $B_4C$ film can be deposited with a similar microstructure as the ones with above 90 % of tabulated bulk density despite of high surface roughness. The $B_4C$ films in this works also show in general good adhesion to the materials and the surface finishing processes (e.g. etching, grinding) commonly used for building neutron detectors.

For the application of neutron converting layers in detectors, a higher density $B_4C$ film means a higher $^{10}B$ density in the film and therefore, more importantly, a higher neutron detection efficiency. The cleanliness of substrates prior and during depositions is an important factor to reduce the deviation in the film properties by reducing the density of surface features and, consequently, the nodular defects. These characteristics on the surface exert clear change on the properties of the $B_4C$ films deposited, which must be taken account of, if important, to a particular designed application.